\documentstyle{article}
\title{
\begin{flushright}
{\bf\normalsize   LPTHE-ORSAY-94-56}\\ \end{flushright}
\bf Frustrating and Diluting \\
Dynamical Lattice Ising Spins
}
\author{{\it D.A. Johnston}\\
         LPTHE\\
         Universite Paris Sud, Batiment 211\\
         F-91405 Orsay, France$^{1}$\\
         }
\date { 7 June 1994 }         
\begin{document}
  \maketitle
                      {\Large
                      \begin{abstract}
%
We investigate
what happens to the third order ferromagnetic phase transition displayed by the
Ising
model on various dynamical planar lattices (ie coupled to 2D
quantum gravity) when we introduce annealed bond
disorder in the form
of either antiferromagnetic couplings
or null couplings. We also look at the effect of such
disordering for the Ising model on general $\phi^3$ and $\phi^4$
Feynman diagrams.
\\ \\
Submitted to Phys Lett B.   \\ \\
$~^{1}$ {\it Address:} Sept. 1993 - 1994, {\it Permanent Address:} Maths Dept,
Heriot-Watt
University, Edinburgh, Scotland \\
%
                        \end{abstract} }
%
  \thispagestyle{empty}
%
%
  \newpage
%
                  \pagenumbering{arabic}
In an earlier paper \cite{0} we observed that the Ising antiferromagnet
would not undergo a phase transition on dynamical $\phi^3$ or $\phi^4$
planar graphs
or on dynamical planar triangulations, essentially because of frustration,
but would display a transition on dynamical planar
quadrangulations. Here we address a related question, namely: what degree of
annealed bond disorder is necessary in order to destroy the ferromagnetic phase
transition
on a dynamical lattice? For simplicity we consider
models in which
there are two possible bond factors, $+J$ appearing with probability $p$
and either $-J$ or $0$ appearing with probability $(1-p)$.
We consider primarily
planar
$\phi^3$ and $\phi^4$ graphs, along with planar triangulations and
quadrangulations but also look more briefly at general $\phi^3$ and
$\phi^4$ graphs of arbitrary topology.
We are thus further complicating the annealed connectivity disorder of the
Ising
model coupled to 2D gravity (ie
living on a planar dynamical graph
or triangulation) by introducing an additional annealed
bond disorder.
Such annealed bond disorder has been considered some years ago for the case
of fixed regular lattices \cite{0a}, where it was shown that the critical
exponents underwent a Fisher renormalization \cite{0b} for models with an
initial \footnote{i.e. In the original
ordered version of the model} $\alpha>0$.
For the two dimensional Ising model it was shown that
introducing a finite fraction of either antiferromagnetic
or null bonds  suppressed
the initial phase transition.
For the Ising model on the dynamical lattices
we consider here with $\alpha = -1$ the considerations in
\cite{0a} will not apply, and it is not immediately obvious what behaviour
to expect.

To investigate the problem we consider the following Ising partition function
on an ensemble of planar random graphs $G^n$ with $n$ vertices.
\begin{equation}
Z_n(\beta J) = \sum_{G^n} \sum_{\{ \sigma \}} \int P (J) dJ \ \exp \left( \beta
J \  \sum_{<ij>} G^n_{ij} \sigma_i \sigma_j \right)
\label{e01}
\end{equation}
where $G^n_{ij}$ is the connectivity matrix for a given graph and
$P(J)$ is the annealed bond distribution.
For the case of a trivial $P(J) = \delta(J -1)$
summing over the number of vertices gives
\begin{equation}
Z(c, g) = \sum_{n=1}^{\infty} \left( { - 4 \lambda c \over (1 - c^2)^2 }
\right)^n Z_n (\beta
J)
\label{e02}
\end{equation}
where $c = \exp ( - 2 \beta)$, so
equ.(\ref{e01}) reduces to the Ising model solved by Boulatov and Kazakov
in \cite{1} by noting the equivalence of equ.(\ref{e02}) to the free energy of
a two matrix model
\begin{equation}
F = { 1 \over N^2} \log \int D^{N^2} \phi_{+} D^{N^2} \phi_{-}  \exp \left( -
{1 \over
2 } \phi_a K_{ab} \phi_b + V(\phi_a, \phi_b) \right)
\label{e03}
\end{equation}
where $a,b = \pm$ and $\phi_{+-}$ are $N \times N$ matrices.
Borrowing the notation of \cite{2a} the inverse propagator $K^{-1}_{ab}$
is given by
\begin{equation}
\begin{array}{cc} K^{-1}_{ab} = & \left(\begin{array}{cc}
\sqrt{g} & { 1 \over \sqrt{g}} \\
{1 \over \sqrt{g}} & \sqrt{g}
\end{array} \right) \end{array}
\end{equation}
where the coupling $g = \exp ( 2 \beta) = 1 / c$
and the potential is
\begin{equation}
V(\phi_+, \phi_-) = {\lambda \over 6 \sqrt{N}} (\phi_+^3 + \phi_-^3 )
\end{equation}
for $\phi^3$ graphs and
\begin{equation}
V(\phi_+, \phi_-) = {\lambda \over 24 N} (\phi_+^4 + \phi_-^4 )
\end{equation}
for $\phi^4$ graphs. In the $\phi^4$ case, after some rescalings,
the matrix model action can be written as
\begin{equation}
S = \phi_+^2 + \phi_-^2 -  2 c \phi_+ \phi_- - {\lambda \over N}
(\phi_+^4 + \phi_-^4)
\label{e06}
\end{equation}
and the matrix integrals in the partition function can be carried
out using the methods of \cite{2} to obtain an exact expression
for the partition function, even in the presence of an external field.
The (non-Onsager) critical exponents that are derived from this expression are
consistent with those obtained in the continuum formulation using
either lightcone or conformal gauge Liouville theory \cite{3}.
In the second paper of \cite{1} the model was also solved on
a $\phi^3$ graphs and the same exponents obtained, thus confirming
universality. This
dynamical lattice universality
is further supported by the results of \cite{4}
for $\phi^3$ graphs without tadpoles and self-energy diagrams,
which again give the same set of critical exponents.
A considerable body of numerical evidence supporting
these results \cite{5} has been accumulated to date.

Let us now consider the effect of introducing a non-trivial
bond distribution $P(J)$ in the above calculations. If we
want to examine the effect of introducing ferromagnetic bonds
with a probability $p$ and antiferromagnetic bonds with a
probability $(1-p)$ we should take $P(J) = p \ \delta (J - 1)
+ (1 - p) \ \delta ( J + 1 )$. This has the effect of modifying
the inverse propagator to
\begin{equation}
\begin{array}{ccc} K^{-1}_{ab} = & p \left(\begin{array}{cc}
\sqrt{g} & { 1 \over \sqrt{g}} \\
{1 \over \sqrt{g}} & \sqrt{g}
\end{array} \right) & + (1-p)
\left(\begin{array}{cc}
{ 1 \over \sqrt{g}} & \sqrt{g}\\
\sqrt{g} & { 1 \over \sqrt{g}}
\end{array} \right)
\end{array}
\end{equation}
We can now invert the propagator \footnote{This is singular for $p=
1/2$, but we run into trouble before reaching this probability.}
and rescale the resulting action to a form resembling that of
equ.(\ref{e06})
\begin{equation}
S = \phi_+^2 + \phi_-^2 - 2 c(p) \phi_+ \phi_- - {\tilde \lambda \over N}
(\phi_+^4 + \phi_-^4)
\label{e07}
\end{equation}
where
\begin{equation}
c(p) = {p  + ( 1 - p ) \ g  \over p \ g + ( 1 -p) }
\end{equation}
and
\begin{equation}
\tilde \lambda = {4 ( g^2 - 1)^2 ( 2 p - 1 )^2
\lambda \over g ( p g + 1 - p )^2
}.
\end{equation}
As we have not changed the form of the action with these manipulations,
merely the interpretation of $c$ and $\lambda$
which are parameters in the action, the
equations for determining the critical point are unchanged. We thus
have, substituting $c(p)$ for $c$ in the appropriate equations in \cite{1},
\begin{equation}
{p  + ( 1 - p ) \ g_c  \over p \ g_c + ( 1 -p) } = {1 \over 4}
\end{equation}
where $g_c$ is the critical value of $\exp 2 \beta$. Solving this
we find
\begin{equation}
\exp 2 \beta_c = {5 p - 1 \over 5p -4}
\end{equation}
which tells us that $\beta_c \rightarrow \infty$ as
$p \rightarrow 0.8$. In other words the critical temperature of the
theory goes to zero as the probability of ferromagnetic bonds drops
to $0.8$ at which point the phase transition vanishes, leaving only
a disordered phase. The value of $p$ at which this happens is
non-universal as can be seen by considering $\phi^3$ lattices. Again the
form of the action is unchanged
by introducing the bond disorder and $c$ in the original ferromagnetic
theory is replaced by $c(p)$ to give the equations
\begin{equation}
{p  + ( 1 - p ) \ g_c  \over p \ g_c + ( 1 -p) } = {\sqrt{28} - 1 \over
27}
\end{equation}
for a pure $\phi^3$ lattice, which gives the vanishing
point $p \simeq 0.86$ and
\begin{equation}
{p  + ( 1 - p ) \ g_c  \over p \ g_c + ( 1 -p) } = {23 \over 108}
\end{equation}
for a $\phi^3$ lattice with no tadpoles and self energies,
which gives the vanishing point $p \simeq 0.82$. It is
perhaps worth remarking that these numbers
are larger than the percolation thresholds
on the appropriate graphs without matter
where they have been calculated
($p \simeq 0.78$ on the
$\phi^3$ graphs with tadpoles and self-energies and
$2/3$ on the $\phi^4$ graphs \cite{6}),
though the back reaction of the matter might be expected to
modify the geometry and hence these values at the Ising
critical points.

We can play similar games in looking at the effects of bond disorder
in bonds of the same sign in the model. If we take, for instance,
$P(J) = p \ \delta(J -1) + (1 - p) \ \delta ( J - 2)$
the propagator becomes
\begin{equation}
\begin{array}{ccc} K^{-1}_{ab} = & p \left(\begin{array}{cc}
 \sqrt{g} & { 1 \over  \sqrt{g}} \\
{1 \over  \sqrt{g}} &  \sqrt{g}
\end{array} \right) & + (1-p)
\left(\begin{array}{cc}
g & { 1 \over g} \\
{1 \over g} &  g
\end{array} \right)
\end{array}
\end{equation}
an we can absorb the change in a
rescaling of $\lambda$ and a $c(p)$ given by
\begin{equation}
c(p) = { p / \sqrt{g} + (1 - p) / g \over p \sqrt{g} + (1 -p) g}.
\end{equation}
On the $\phi^4$ lattice for instance
the critical value of $g$ will thus interpolate smoothly
between $4$ and $2$ as $p$ changes from $1$ to $0$. For
$P(J)$ of the form $p \ \delta(J -1) + (1 - p)  \ \delta ( J - \Delta)$
the interpolation would be between $4$ and $4^{(1/\Delta)}$.

The case of bond dilution, in which null bonds are inserted with probability
$(1-p)$ is particularly intriguing. We have $P(J) = p \ \delta(J -1) + (1 - p)
\ \delta (
J)$ in this case, so
\begin{equation}
\begin{array}{ccc} K^{-1}_{ab} = & p \left(\begin{array}{cc}
 \sqrt{g} & { 1 \over  \sqrt{g}} \\
{1 \over  \sqrt{g}} &  \sqrt{g}
\end{array} \right) & + (1-p)
\left(\begin{array}{cc}
1 &  1  \\
1 &  1
\end{array} \right)
\end{array}
\end{equation}
which in turn leads to a $c(p)$ of the form
\begin{equation}
c(p) = { p / \sqrt{g} + (1-p) \over p \sqrt{g} + (1 - p)}
\end{equation}
and
\begin{equation}
\tilde \lambda ={ 4 p^2 ( g - 1 )^2 ( p (g + 1)
 + 2 ( 1 - p) \sqrt{g} )^2 \lambda
\over ( p \sqrt{g} + 1 - p )^2 g^2 }.
\end{equation}
If we now solve the equation for the critical point $c(p) = 1/4$ on $\phi^4$
graphs,
we find that
\begin{equation}
\exp ( \beta_c ) = { 3 ( 1 - p) + \sqrt{ 9 ( 1 - p)^2 + 16 p^2} \over 2 p }
\end{equation}
which gives a $\beta_c \rightarrow \infty$ only as $p \rightarrow 0$.
A similar result applies for the bond dilute model on $\phi^3$ graphs.
The critical temperature goes to zero and the transition vanishes
only when the probability of non-null bonds goes to zero.
This is in direct contrast to the case of annealed disorder with dilute
bonds on a fixed lattice where the transition vanishes for a finite
value of $p$ that is lower than in the case of competing bond signs
and close to the percolation threshold for the lattices \cite{0a}.
It is rather surprising that the transition should persist on a dynamical
lattice down to zero active bond concentration - perhaps the dynamical
nature of the lattice allows sufficient communication between the active bonds
to generate a transition by some sort of clustering mechanism. In the fixed
lattice
case small, but definite, correlations are present between like bonds, so such
an
effect could well be magnified on any sort of dynamical lattice.

In \cite{0} we observed that it was also possible to write
down directly matrix models for Ising spins on dynamical
triangulations
\begin{equation}
S =  { N \over g} tr \left( {1 \over 2 \cosh (\beta) (1 + c^*)} S^2 + { 1 \over
2 \cosh
( \beta) ( 1 - c^*)} D^2
+ S^3 / 3 + S D^2 \right)
\end{equation}
where $S$ represented edges of triangles with the same spins at each end
and $D$ edges of triangles with different spins. The coupling $c^*$ was the
``dual'' to $c$, $(1 - c) / (1 + c)$. The model could be rescaled into the
$O(1)$ representation of the Ising model written down in \cite{8}
and gave the correct dual critical temperature. We can generalize the
arguments of \cite{0} to an action of the form
\begin{equation}
S =  { N \over g} tr \left( {1 \over 2}  A(\beta) S^2 + { 1 \over 2 } B (\beta)
D^2
+ S^3 / 3 + S D^2 \right)
\end{equation}
which, using the results of \cite{8}, will display Ising critical behaviour if
$A$ and $B$ satisfy
\begin{equation}
7 B(\beta)^2 + 6 A (\beta)^2 - 14 A ( \beta) B ( \beta) =0.
\label{e10}
\end{equation}
The propagator is diagonal in such models, which makes it particularly easy to
write
down the effect of non-trivial $P(J)$s. For $P(J) = p \ \delta(J - 1) + ( 1 -
p) \
\delta (J + 1)$ we have $1/A(\beta,p) = \cosh (\beta) ( 1 + (2 p - 1) c^*)$ and
$1/B(\beta,p) = \cosh (\beta) ( 1 - (2 p - 1) c^*)$. Dividing through by
$A^2$ in equ.(\ref{e10}) we see that Ising critical behaviour is
possible if the ratio $A/B$ takes the same value as in the case $p=1$,
namely $(14 - \sqrt{7})/(13 + \sqrt{7})$.
We thus find that
\begin{equation}
c^* = { 1 \over 2 p - 1} { \sqrt{28} - 1 \over 27}
\end{equation}
which gives a minimum value of $p$, since we know that a ferromagnetic
transition must
have $c^* < 1$. The transition is pushed to zero temperature ($c^* = 1$) at
$p \simeq 0.58$, leaving only a disordered phase.
This is
entirely analogous with the model on $\phi^3$ and $\phi^4$ graphs.
Taking the ratio
of the couplings in the bond dilute case
where $1/A(\beta,p) = p \cosh (\beta) ( 1 + c^* ) + ( 1 - p)$ and
$1/B(\beta,p) = p \cosh (\beta) ( 1 - c^*) + (1 - p)$, gives the following
equation
for the critical temperature
\begin{equation}
 \cosh(\beta_c) - 6.291 \sinh(\beta_c) \simeq   ( p - 1) /p.
\end{equation}
This has solutions for arbitrarily small $p$ so, just as for the dilute bond
Ising model on planar $\phi^3$ and
$\phi^4$ graphs, the phase transition only vanishes as $p \rightarrow 0$.

The matrix model for dynamical quadrangulations considered in \cite{0},
which can be written in a rescaled form as
\begin{equation}
S = {N \over g} tr \left( {1 \over 2 \cosh (\beta) (1 + c^*)} S^2  + { 1 \over
2 \cosh
(\beta) (1 - c^*)} D^2 + {1 \over
4 } S^4 + {1\over 4 } D^4
 + {1 \over 2 } ( S D S D + 2 S^2 D^2 ) \right),
\end{equation}
can also
be modified in a similar fashion to
the triangulation model to include bond disorder:
\begin{equation}
S = {N \over g} tr \left( {1 \over 2  } A(\beta) S^2  + { 1 \over 2 }
B (\beta)  D^2 + {1 \over
4 } S^4 + {1\over 4 } D^4
 + {1 \over 2 } ( S D S D + 2 S^2 D^2 ) \right).
\end{equation}
If we include antiferromagnetic bonds,
$P(J) = p \ \delta(J+1) + (1 - p) \ \delta(J -
1)$,
and demand that the ratio $B/A$ be the same as in
the original model we find
\begin{equation}
c^* = { 1 \over 2 p - 1} { 1 \over 4}.
\end{equation}
Noting that $c^*$ should be less than $1$ again
gives a critical value of $p=5/8$
for the vanishing of the ferromagnetic transition.
The dynamical quadrangulation
model will also have an antiferromagnetic transition, which will
appear for $p<3/8$ by similar arguments
upon replacing $c^*$ by $-c^*$ (ie $\beta$ by $ - \beta$).
For $3/8<p<5/8$ there would appear to be no
transition at all.
The bond dilute model gives the following equation for the critical temperature
\begin{equation}
\cosh(\beta_c) - 4 \sinh(\beta_c) =   ( p - 1) /p
\end{equation}
so the critical temperature will only go to zero as $p \rightarrow 0$.

The planar $\phi^3$ and $\phi^4$ graphs and
their duals that we have looked at here can
be considered as an ensemble of ``fat'' or ribbon graphs
generated by the perturbative $N \times N$ matrix integrals
in the limit $N \rightarrow \infty$, with the spin
models living on them. In a recent paper \cite{2a} the opposite limit
of $N \rightarrow 1$ was considered, which gives an ensemble of
standard ``thin'' Feynman diagrams. In
an annealed ensemble of such graphs, which are locally
tree like, the Ising model displays a mean field transition at the
Bethe lattice values of $g_c$, namely $3$ for $\phi^3$ graphs and $2$
for $\phi^4$ graphs. Introducing a coupling distribution of the form
$P(J) = p \ \delta ( J - 1 ) + (1 - p ) \ \delta (J + 1)$ on
thin $\phi^3$ graphs
and scaling the action the form
\begin{equation}
S = { 1 \over 2} (\phi_+^2 + \phi_-^2) - c(p) \phi_+ \phi_-  - {1 \over 3}
(\phi_+^3 +
\phi_-^3)
\end{equation}
gives the following saddle point values for
$\phi_+, \phi_-$
\begin{eqnarray}
\phi_+,\phi_- &=&  { ( g - 1 ) \, p \over
        ( 1 - p ) \, \sqrt{g} + p g}  \nonumber \\
\phi_+,\phi_- &=& { 1 + g \pm \sqrt{ (g + 1) ( g ( 4 p - 3 ) - (4 p -1))}
\over 2 ( ( 1- p ) + p g ) }
\end{eqnarray}
(along with a trivial zero solution and a solution with $\phi_+$ and
$\phi_-$ interchanged) \cite{2a}. The low temperature ordered
phase is given by the second solution with differing $\phi_+,\phi_-$, so
this is only possible when the expression inside the square root is
positive. The critical $g_c$ is that for which the square root is
zero, namely
\begin{equation}
g_c = { 4 p - 1 \over 4 p - 3}.
\end{equation}
Once again the critical temperature goes to
zero at a finite value of $p$, leaving only the disordered
phase for $p<3/4$. The rather peculiar behaviour of the dilute
model with $P(J) = p \ \delta(J - 1) + (1-p) \ \delta(J)$
also appears to persist on thin graphs, where we have the following
saddle point solutions for $\phi_+, \phi_-$
\begin{eqnarray}
\phi_+,\phi_- &=&  { ( g - 1 ) \, p \over
        ( 1 - p ) \, \sqrt{g} + p g}  \nonumber\\
\phi_+,\phi_- &=&  { 2  \sqrt{g} ( 1 - p ) + p ( g + 1) \pm
         \sqrt{p^2 g^2 - 2( 3p^2 - 4p +2) g - 8p(1-p) \sqrt{g} - 3p^2}
        \over 2 ( \sqrt{g} ( 1 - p) + g p )}
\end{eqnarray}
In this case we find that it {\it is} possible to obtain a positive
term inside the square root for all values of $p$, with the critical
temperature going to zero only as $p \rightarrow 0$.

The saddle point equations for general $\phi^4$ graphs may be derived in a
similar fashion from the action
\begin{equation}
S = { 1 \over 2} (\phi_+^2 + \phi_-^2) - c(p) \phi_+ \phi_-  - {1 \over 4}
(\phi_+^4 + \phi_-^4)
\end{equation}
for the various $c(p)$. For purely
ferromagnetic interactions,$c(p) = 1 / g$, the
high and low temperature saddle point values are
\begin{eqnarray}
\phi_+,\phi_- &=& - \sqrt{ 1 - 1 / g } \nonumber\\
\phi_+ &=& { \left( \sqrt{ 1 - \sqrt{1 - 4 / g^2}}\right)
\left( 1 + \sqrt{1 - 4 / g^2} \right) g \over  2 \sqrt{2}},
\nonumber\\
\phi_- &=& { \sqrt{ 1 - \sqrt{1 - 4 / g^2}} \over \sqrt{2}}.
\end{eqnarray}
We find that for antiferromagnetic bonds,
$P(J) = p \ \delta(J - 1) + (1-p) \ \delta(J + 1)$,
the ordered phase is only possible for $p>2/3$, with
the critical value of $g$ being given by
\begin{equation}
g_c = {3 p - 1  \over 3 p - 2}.
\end{equation}
Bond dilution gives a transition that
persists for all finite $p$ in this case too.

In summary, we have investigated the effects of including annealed
bond disorder and dilution on the Ising model in
ensembles of planar
$\phi^3$ and $\phi^4$ graphs, triangulations,
quadrangulations and general $\phi^3$ and $\phi^4$ graphs. We find that a fixed
finite probability of antiferromagnetic bonds is sufficient to
suppress the Ising transition in all cases, but that the
transition persists down to zero active bond probability in the
case of bond dilution, with the transition temperature going to zero
as $p \rightarrow 0$.
It would be interesting to attempt to understand
the persistence of the bond dilute transition by looking at
bond/bond correlations in the various models in a similar fashion
to the fixed lattice work in \cite{0a}.
It might also be possible to extend the calculations to
bond distributions other than the sums of delta functions considered here.
{}From the numerical point of view it would be an amusing exercise to
perform simulations to
attempt to verify the predictions for
the critical concentrations with antiferromagnetic bonds
and to see if the transitions in the bond dilute models really are
as tenacious as the calculations here suggest.

We have not considered {\it quenched}
connectivity or bond disorder at all in this paper.
There is no problem in dealing with this on the
thin graphs and the calculations in \cite{2a} indicate that
interesting effects such as a spin glass phase may appear on the introduction
of quenched bond disorder - or simply antiferromagnetic couplings.
For ``fat'' graphs we immediately run into problems with the $c=1$
barrier when we employ the replica trick to calculate
logarithms of partition functions. For the Ising model we can only
caculate with $n=1,2$ replicas, for instance
\footnote{If we ignore the problem, naively taking $n \rightarrow 0$
in the $KPZ/DDK$ formulae gives exponents for the Ising and Potts models that
are
close to the dynamical lattice ones. Amusingly, a simulation on a quenched
ensemble of
planar $\phi^3$ graphs does give exponents that are close
to those on a dynamical lattice \cite{9}.}.
In view of the importance of the replica trick in this domain
any insights on matrix models for $c>1$ are likely to have some bearing
on the treatment of quenched disorder.

D. Johnston was supported at Orsay by an EEC Human Capital and Mobility
Fellowship and an Alliance grant.

\vfill
\eject

\bigskip
\begin{thebibliography}{99}
\bibitem{0} D. Johnston, Phys. Lett. {\bf B314} (1993) 69.
\bibitem{0a} M. Thorpe and D. Beeman, Phys. Rev. {\bf B14} (1976) 188.
\bibitem{0b} M. Fisher, Phys. Rev. {\bf 176} (1968) 257.
\bibitem{1} V. A. Kazakov, Phys. Lett. {\bf A119} (1986) 140.\\
             D.V. Boulatov and V.A. Kazakov, Phys. Lett. {\bf B186} (1987) 379.
\bibitem{2a} C. Bachas, C. de Calan and P. Petropoulos, ``Quenched Random
Graphs'', Ecole Polytechnique Preprint CPTH-A264.1093
(draft version), hep-th 9405068.
\bibitem{2} E. Brezin, C. Itzykson, G. Parisi and J.B. Zuber, Commun. Math.
Phys. {\bf 59} (1978) 35\\
            M.L. Mehta, Commun. Math. Phys. {\bf 79} (1981) 327.
\bibitem{3} V.G. Knizhnik, A.M. Polyakov and A.B. Zamolodchikov, Mod. Phys.
Lett. {\bf A3} (1988) 819.\\
            F. David, Mod. Phys. Lett. {\bf A3} (1988) 1651\\
            J. Distler and H. Kawai, Nucl. Phys. {\bf B321} (1989) 509.
\bibitem{4} Z. Burda and J. Jurkiewicz, Acta Physica Polonica {\bf B20} (1989)
949.
\bibitem{5} J. Jurkiewicz, A. Krzywicki, B. Petersson and B. Soderberg, Phys.
Lett. {\bf B213} (1988) 511.\\
            R. Ben-Av, J. Kinar and S. Solomon, Nucl. Phys. {\bf B ( Proc.
Suppl.) 20} (1991) 711.\\
            S.M. Catterall, J.B. Kogut and R.L. Renken, Phys. Rev. {\bf D45}
(1992) 2957\\
            C. Baillie and D. Johnston,  Mod. Phys. Lett. {\bf A7} (1992)
1519.\\
            J. Ambj\o rn, B. Durhuus, T. Jonsson and  G. Thorleifsson, Nucl.
 Phys. {\bf B398} (1992) 568.
\bibitem{6} V. Kazakov, Mod. Phys. Lett. {\bf A 17} (1989) 1691.
\bibitem{8} B. Eynard and J. Zinn-Justin, Nucl. Phys. {\bf B386} (1992) 558.
\bibitem{9} C. Baillie, K. Hawick and D. Johnston, Phys. Lett. {\bf B328}
(1994) 251.
\end{thebibliography}
\end{document}